\documentclass[prd,showpacs,showkeys,amsmath,amssymb,10pt]{revtex4}
\usepackage{amsfonts}

\usepackage{graphicx}
\usepackage{dcolumn}
\usepackage{bm}
\def\ut#1{\rlap{\lower1ex\hbox{$\sim$}}{#1}}

\begin{document}

\title{Topological Gravity Localization on a $\delta$-function like Brane}
\author{M. O. Tahim}
\email{mktahim@fisica.ufc.br}
\author{C. A. S. Almeida}
\email{carlos@fisica.ufc.br} \affiliation{Departamento de
F\'{\i}sica, Universidade Federal do Cear\'a, Caixa Postal 6030,
60455-760, Fortaleza, Cear\'a, Brazil}

\begin{abstract}
Besides the String Theory context, the quantum General Relativity
can be studied by the use of constrained topological field theories.
In the celebrated Plebanski formalism, the constraints connecting
topological field theories and gravity are imposed in space-times
with trivial topology. In the braneworld context there are two
distinct regions of the space-time, namely, the bulk and the
braneworld volume. In this work we show how to construct topological
gravity in a scenario containing one extra dimension and a
$\delta$-function like $3$-brane which naturally emerges from a
spontaneously broken discrete symmetry. Starting from a $D=5$ theory
we obtain the action for General Relativity in the Palatini form in
the bulk as well as in the braneworld volume. This result is
important for future insights about quantum gravity in brane
scenarios.

\end{abstract}
\maketitle

\maketitle



Physics of extra dimensions has been successful. Indeed, some of the
crucial drawbacks of the Standard Model are better understood when
the theories are embedded in $D>4$ space-times. In the context of
String Theory and Branes, we have at hand a rather general and
simple view of how a theory of grand unification should be
constructed \cite{polchinski}. In Braneworld models, the Standard
Model is completely embedded in scenarios containing extra
dimensions and intersecting branes \cite{braneworld}. In particular,
the Randall-Sundrum scenario \cite{RS} gives an elegant solution to
an old problem of the Standard Model: the gauge hierarchy problem.
Besides, we can understand how the gravitational field can be
trapped to the $4$-dimensional world. Extensions of the
Randall-Sundrum model in order to study supersymmetry \cite{susy}
have been made, including in the context of String Theory
\cite{stringextraD}.

The results of the Randall-Sundrum model are based in the classical
theory of General Relativity, a theory that possesses as a
fundamental quantity the space-time metric $g_{\mu\nu}$. However, as
is well known, the Einstein-Hilbert formulation of gravitation does
not support the common procedures of quantization. In other words,
the quantum theory of gravity suffer problems of
non-renormalizability. In order to avoid this bad characteristic new
methods were developed \cite{loop}. Such approaches are based on
topological gravity theories, i. e., theories where the metric
$g_{\mu\nu}$ is derived from more fundamental quantities. In the
context of stringy models, the idea of background independence is
very important in order to fully understand the dynamics of
space-time \cite{backgroundwitten}.

In view of these developments, it is interesting to build and study
topological gravity models in the context of extra dimensions and
branes. In this direction, some discussions have already been made.
Boyarsky and Kulik \cite{boyarsky} have studied topological gravity
on a singular boundary brane from a higher derivative topological
theory in order to link their model with solitonic brane
backgrounds. On the other hand a BF topological theory which induces
gravity on a boundary through specific conditions was discussed by
Henty \cite{henty}. In particular recently we have proposed an
approach to the gauge hierarchy problem from the topological gravity
viewpoint \cite{ourwork}. In that case, no mention is made about
gravity in the bulk $D=5$ space-time.

In this letter, we show how to construct topological gravity in a
scenario containing one infinity extra dimension and a
$\delta$-function like brane. In the Plebanski formalism, the link
between a Topological Field Theory and General Relativity is made
via constraints expressed by Lagrange multiplier fields $\ut{\Phi}$.
The important detail is that the constraints are valid for a
space-time without any mention to a boundary. In our case, we have
two distinct regions of the space-time: the $D=5$ bulk and the $D=4$
domain hyperplane ($3$-brane). Where should we implement the
constraints? The answer is that if we impose the constraints in
$D=5$ bulk, then we can build up the Palatini action in the bulk
space-time and also in the $3$-brane after a field identification.

Topological gravity can be studied by means of constrained
topological field theories. Here we follow the conventions
introduced by Freidel \textit{et al.} \cite{freidel}. Thus, we
regard Greek characters as space-time indices, Latin letters are
internal indices, and a tilde over the fields represent the fact
that they are tensorial densities. Our model is based in the
following action in five dimensions:
\begin{equation}\label{um}
S=\int dx^{5} \left[\widetilde{B}^{\mu\nu}_{ij} F^{ij}_{\mu\nu}
+\frac{1}{2} \ut{\Phi}_{m\mu\nu\rho\sigma}\epsilon^{mijkl}
\widetilde{B}^{\mu\nu}_{ij}\widetilde{B}^{\rho\sigma}_{kl}+
L_{brane}+ k \partial_{\alpha}\varphi
\widetilde{C}^{\mu\nu\alpha}_{ij} F^{ij}_{\mu\nu}\right],
\end{equation}
where
$L_{brane}=\frac{1}{2}\partial_{\mu}\varphi\partial^{\mu}\varphi
-V(\varphi)$. This action is a functional of SO$(4,1)$ gauge
fields $A^{ij}_{\mu}$, bivector fields
$\widetilde{B}^{\mu\nu}_{ij}$, the fields
$\widetilde{C}^{\mu\nu\alpha}_{ij}$, Lagrange multiplier fields
$\ut{\Phi}_{m\mu\nu\rho\sigma}$ and a real scalar field $\varphi$.
The two first terms define the Plebanski action in $D=5$. The
second and third terms represent the part of the action that
generates the $3$-brane of this model. Indeed, this $3$-brane is a
domain wall embedded in a $D=5$ space-time. For this, we suppose
$\varphi=\varphi(x_{4})$ and use
$V(\varphi)=\lambda(1-\cos\varphi)$. The last term in
Eq.(\ref{um}) is a topological term that will give an effective BF
type action over the $3$-brane in $D=4$. It is just this term the
responsible for topological gravity on the brane.

Having made the first presentations, it is now important to
discuss the symmetries of this model. The first one is a discrete
symmetry: the action is invariant under the change
$\varphi\rightarrow \varphi+2\pi$ (it is like a Peccei-Quinn
symmetry). This symmetry is important because, if spontaneously
broken, it will give us kink like defects. The defect related to
the topological sector containing just one soliton will be the
$3$-brane of the model (the scalar field model treated here is the
sine-Gordon one that, as it is well known, has several solitonic
solutions). The second type of symmetry is the general coordinate
transformation. In this case, this symmetry plays the role of a
SO$(4,1)$ gauge symmetry. Despite the terms responsible for the
brane, the action is generally covariant: the fields
$\widetilde{B}^{\mu\nu}_{ij}$ and
$\widetilde{C}^{\mu\nu\alpha}_{ij}$ scales as tensor densities of
weight one, while the multipliers scale as tensorial densities of
weight minus one (represented as a single tilde below the symbol
`$\Phi$`). The role played by the $3$-brane it is to break this
SO$(4,1)$ symmetry in order to induce an SO$(3,1)$ symmetry over
itself. This fact is not a surprise. In string theories, for
instance, this is a basic characteristic of models containing
D-branes.

Before study the main idea of this work, we review quickly some
aspects of the Plebanski formulation of topological gravity. More
rigorous details can be found in Ref. \cite{freidel} and
references therein. The fields $\widetilde{B}^{\mu\nu}_{ij}$ and
$\widetilde{C}^{\mu\nu\alpha}_{ij}$ are defined in the following
manner:
\begin{equation}\label{dois}
\widetilde{B}^{\mu\nu}_{ij}=\frac{1}{4}\widetilde{\epsilon}^{\mu\nu\alpha\beta\lambda}
B_{\alpha\beta\lambda  ij}
\end{equation}
and
\begin{equation}\label{tres}
\widetilde{C}^{\mu\nu\alpha}_{ij}=\frac{1}{12}\widetilde{\epsilon}^{\mu\nu\alpha\beta\lambda}
C_{\beta\lambda  ij}.
\end{equation}

In order to write an action for gravity we must compute the
variation of the action (\ref{um}) with the Lagrange multiplier
field $\ut{\Phi}$. For such, we have to postulate the following
property obeyed by the Lagrange multiplier:
\begin{equation}\label{quatro}
\epsilon^{m\mu\nu\rho\sigma}\ut{\Phi}_{m\mu\nu\rho\sigma}=0.
\end{equation}
Therefore, the variation of the action (\ref{um}) give us
\begin{equation}\label{cinco}
\frac{\delta\ut{\Phi}_{m\mu\nu\rho\sigma}}{\delta\ut{\Phi}_{n\alpha\beta\gamma\lambda}}\epsilon^{mijkl}
\widetilde{B}^{\mu\nu}_{ij} \widetilde{B}^{\rho\sigma}_{kl}=0.
\end{equation}
Comparing the variation of Eq. (\ref{quatro}) with the Eq.
(\ref{cinco}) we obtain, for some coefficients $c^{m}_{\alpha}$,
that
\begin{equation}\label{seis}
\epsilon^{mijkl} \widetilde{B}^{\mu\nu}_{ij}
\widetilde{B}^{\rho\sigma}_{kl}=c^{m}_{\alpha}\epsilon^{\alpha\mu\nu\rho\sigma}.
\end{equation}
The coefficients $c^{m}_{\alpha}$ then satisfy:
\begin{equation}\label{sete}
c^{m}_{\alpha}=\frac{1}{5!}\epsilon_{\alpha\mu\nu\rho\sigma}\epsilon^{mijkl}\widetilde{B}^{\mu\nu}_{ij}
\widetilde{B}^{\rho\sigma}_{kl}.
\end{equation}
The major importance of Eq. (\ref{sete}) is viewed in the following
theorem introduced by Feidel \textit{et. al.} \cite{freidel}:

*\underline{Theorem}: In $D>4$, a generic B field satisfies the
constraints of Eq. (\ref{sete}) if and only if it comes from a frame
field. In another words, if B is non-degenerate and satisfies Eq.
(\ref{sete}), then there are frame fields $e^{\mu}_{i}$ such that
\begin{equation}\label{oito}
\widetilde{B}^{\mu\nu}_{ij}=\pm |e| e^{[\mu}_{i}e^{\nu]}_{j}.
\end{equation}
In the last equation, $|e|$ is the determinant of the frame field
$e^{\mu}_{i}$. Now, is important to note that the constraints of Eq.
(\ref{sete}) give $5$-dimensional $e^{\mu}_{i}$ fields, i.e., we are
imposing constraints in the bulk space-time. Let us see what are the
consequences for the physics on the brane. We must consider that in
the vicinity of the brane worldsheet the metric may be written in
terms of coordinates $\xi^a$ and $w$ ($w$ is the normal distance
from the hyperplane) in the following way:
\begin{equation}\label{setenta}
d s^2=\gamma_{ab}d\xi^{a}d\xi^{b}-d w^2.
\end{equation}
In the last equation
$\gamma_{ab}=g_{\mu\nu}x^{\mu}_{,a}x^{\nu}_{,b}$ is the worldsheet
induced metric (this metric is constructed with the frame fields
encountered in the theorem above). The equation of motion for the
$\varphi$ field, regarded as static and having only dependence in
the $x^5\equiv w$ coordinate ($\varphi=\varphi(w)$), is given by:
\begin{equation}\label{nove}
-\frac{d^{2}\varphi}{d
w^{2}}+\frac{dV(\varphi)}{d\varphi}-\frac{d}{d\varphi}(k
\widetilde{C}^{\mu\nu 4}_{ij} F^{ij}_{\mu\nu})=0.
\end{equation}
Putting $V(\varphi)=\lambda (1-\cos\varphi)$ and discarding
fluctuations due to the fields $\widetilde{C}^{\mu\nu\alpha}_{ij}$
and $F^{ij}_{\mu\nu}$ we conclude that
\begin{equation}\label{dez}
-\frac{d^{2}\varphi}{d w^{2}}+\lambda\sin\varphi=0,
\end{equation}
which solution is the solitonic $3$-brane of this model (in this
step, the symmetry $\varphi\rightarrow \varphi+2\pi$ is
spontaneously broken, favoring the appearance of domain
hyperplanes), namely:
\begin{equation}\label{onze}
\varphi(w)=4 tg^{-1} exp \left(\sqrt{\lambda}w\right)
\end{equation}
In contrast with present works in topological gravity where the
brane is introduced \textit{ad hoc}, it is worth mentioning that
here the $3$-brane appears from a very natural physical way.

Let us return to the action (\ref{um}) and see its last term, i.e.,
\begin{equation}\label{doze}
S_{brane}\sim k\int d^{5}x \left[\frac{d\varphi(w)}{dw}
\widetilde{C}^{\mu\nu 4}_{ij} F^{ij}_{\mu\nu}\right].
\end{equation}
In the thin hyperplane limit we have $\frac{d\varphi(w)}{dw}\sim
\delta(w)$. Then, from Eq. (\ref{doze}) we obtain an effectively
$4$-dimensional BF like action:
\begin{equation}\label{treze}
S_{4d}\mapsto k\int d^{4}x \left[ \widetilde{C}^{\mu\nu 4}_{ij}
F^{ij}_{\mu\nu}\right].
\end{equation}
The interpretation is that, regarding the conditions discussed, some
sort of BF theory should describe the physics in the $4$-dimensional
world. It is convenient to note that the breaking of the discrete
symmetry of this model induces the explicit break of the SO$(4,1)$
gauge symmetry down to a SO$(3,1)$ symmetry in the brane. The Greek
and Latin indices are enumerated from $0$ till $4$ in $D=5$. By
construction (see Eq. (\ref{tres})), if we are studying a
$4$-dimensional world, the indices match. Now, the most important
thing. If we identify the degrees of freedom of
$\widetilde{C}^{\mu\nu 4}_{ij}$ with that coming from Eq.
(\ref{oito}), i.e., if
\begin{equation}\label{catorze}
\widetilde{C}^{\mu\nu 4}_{ij}|_{D=4}\equiv
\widetilde{B}^{\mu\nu}_{ij}=\pm |e^{(5)}| e^{[\mu}_{i}e^{\nu]}_{j}
|_{D=5},
\end{equation}
then we can substitute this result back in the original action
(Eq. (\ref{um})) and rewrite it effectively as
\begin{equation}\label{quinze}
S_{eff.}=\pm \int d^{5}x \left[|e^{(5)}| e^{[M}_{I}e^{N]}_{J}
F^{IJ}_{MN}+k'\delta(w) |e^{(4)}|
e^{[\mu}_{i}e^{\nu]}_{j}F^{ij}_{\mu\nu}\right].
\end{equation}
Here we have written capital Latin indices in order to make
distinction between $D=5$ and $D=4$ internal and space-time
indices. The result is just the Palatini action for $D=5$ gravity
together a new term. This new term is the Palatini action for
gravity in a $3$-brane of the $\delta$-function type. The
resemblance between this result and that obtained by Dvali
\textit{et. al.} \cite{Dvali} should be noted. In that work the
action (\ref{quinze}) can be obtained by loop corrections due to
localized matter on the brane or due to specific couplings between
gravity and scalar fields. Finally, a simple detail should be
considered. The identification procedure established by Eq.
(\ref{catorze}) depends specifically on $|e^{(5)}|$. Fortunately,
a $4$-dimensional determinant may be constructed from it. Here, we
must consider, for instance, that the new constant $k'$ in Eq.
(\ref{quinze}) depends on the extra dimension. This indeed may be
of great importance in order to understand the physical effects of
the extra dimension in $D=4$ in the context of topological
gravity.

The main conclusion of this work is that we can construct a
$\delta$-like brane scenario in the context of topological gravity
by the use of a topological constrained field theory. On the other
hand, we use a spontaneously breaking symmetry mechanism in order to
generate the $3$-brane, which it is a new approach in this context.
The procedure introduced here depends exclusively on constraints
imposed on the bulk $D=5$ space-time. Then, we obtain gravity theory
in $D=4$ starting from a very specific interaction in the action
(\ref{um}). In discussions about the \textit{physical reality} of
the formalism of topological gravity, the non-dynamical nature of
the degrees of freedom in this kind of theory is often cited. In
this regards, the model described here looks like some others
gravity localization approaches. In these approaches, some
\textit{physical} gravitational degrees of freedom are effectively
trapped to a $3$-brane. However, in the context of topological
gravity, it is hard to affirm if such a mechanism could happen or
not. The unique characteristics that guarantee the existence of a
$4$-dimensional theory in the brane are mathematical: SO$(3,1)$
gauge symmetry $\Rightarrow D=4$ theory.

An interesting question is the cosmological constant. A
generalization of the Plebanski action results in the Palatini
action for general relativity plus a cosmological constant term
\cite{palatinicosmological}. In the lines discussed here, the
generation, if possible, of cosmological constant in the brane
should be investigated. In the context of topological gravity,
another interesting subject is the Immirzi parameter \cite{immirzi}
and the related consequences within a brane formalism.

A natural further step would be an attempt to build up a topological
version of the Randall-Sundrum scenario \cite{RS}. The motivations
behind this step are several: the possibility of quantization of the
scenario; alternatives to solve the standard model problems such as
the hierarchy problem, etc.

The importance of our result lies in the fact that it is possible
to fully quantize topological gravity models. To the best of our
knowledge, physical consequences related to the existence of extra
dimensions have been well formulated in the context of classical
theories, not in the context of quantum models. The quantum
consequences are welcome and interesting to be studied because of
at least two good reasons: i) important and modern results (like
that of the Randall-Sundrum scenario) could be reinforced and ii)
we may or may not decide in favor of the formalisms of topological
gravity. We believe that our result should help in future attempts
to quantize gravity in brane backgrounds.

The authors would like to thank Funda\c{c}\~{a}o Cearense de apoio
ao Desenvolvimento Cient\'{\i}fico e Tecnol\'{o}gico (FUNCAP) and
Conselho Nacional de Desenvolvimento Cient\'{\i}fico e
Tecnol\'{o}gico (CNPq) for financial support.

\end{document}